\def\ra{\rangle}
\def\la{\langle}
\def\be{\begin{equation}}
\def\ee{\end{equation}}
\def\ba{\begin{array}}
\def\ea{\end{array}}

\def\Cb{{\Bbb C}}

\documentstyle[12pt]{article}
\input amssym.def
\begin{document}
\baselineskip=18pt \setcounter{page}{1} \centerline{\large\bf
Lower Bound of Concurrence for Multipartite Quantum States}
\vspace{4ex}
\begin{center}
Ming Li$^{1}$, Shao-Ming Fei$^{1,2}$ and  Zhi-Xi Wang$^{1}$

\vspace{2ex}

\begin{minipage}{5in}

\small $~^{1}$ {\small Department of Mathematics, Capital Normal
University, Beijing 100037}

{\small $~^{2}$ Institut f\"ur Angewandte Mathematik, Universit\"at
Bonn, D-53115}


\end{minipage}
\end{center}

\begin{center}
\begin{minipage}{5in}
\vspace{1ex} \centerline{\large Abstract} \vspace{1ex}
We study the concurrence of arbitrary multipartite mixed quantum states.
An explicit lower bound of the concurrence is derived, which detects quantum
entanglement of some states better than some separability criteria,
and gives sufficient conditions for distilling GHZ states from tripartite states.
An interesting relations between the lower bound of the concurrence
for bipartite states and for tripartite states has been presented.

\smallskip
PACS numbers: 03.67.-a, 02.20.Hj, 03.65.-w\vfill
\smallskip
\end{minipage}\end{center}
\bigskip

\section{\bf Introduction}
Quantum entanglement plays crucial roles in quantum information
processing \cite{nielsen}. Entanglement of formation (EOF)
\cite{eof} and concurrence \cite{concurrence,anote} are two well
defined quantitative measures of quantum entanglement. For two-quibt
systems it has been proved that EOF is a monotonically increasing
function of the concurrence and an elegant formula for the
concurrence was derived analytically by Wootters \cite{wotters}.
However with the increasing dimensions of the subsystems the
computation of EOF and concurrence become formidably difficult. A
few explicit analytic formulae for EOF and concurrence have been
found only for some special symmetric states
\cite{Terhal-Voll2000,fjlw,fl,fwz,Rungta03}.

The first analytic lower bound of concurrence
that can be tightened by numerical optimization over some parameters
was derived in \cite{167902}.
In \cite{Chen-Albeverio-Fei1,chen} analytic
lower bounds on EOF and concurrence for any dimensional mixed
bipartite quantum states have been presented
by using the positive partial
transposition (PPT) and realignment separability criteria.
These bounds are exact for some
special classes of states and can be used to detect many bound
entangled states. In \cite{breuer} another lower bound on EOF for
bipartite states has been presented from a new separability
criterion \cite{breuerprl}. A lower bound of concurrence based on
local uncertainty relations (LURs) criterion is derived in
\cite{vicente}. This bound is further optimized in \cite{zhang}. The
lower bound of concurrence for tripartite systems has been studied
in \cite{gao}.

In \cite{edward,ou} the authors presented
lower bounds of concurrence for bipartite systems in terms of
a different approach. It has been shown that this lower bound has a
close relationship with the distillability of bipartite quantum states.

In this letter, we study the lower bound of concurrence for
arbitrary multipartite quantum systems by using the approach in
\cite{ou}. Let $H$ denotes a $d$-dimensional vector space with basis
$|i\ra$, $i=1,2,...,d$. An $N$-partite pure state in
${H}\otimes\cdots\otimes{H}$ is generally of the form,
\begin{eqnarray}\label{purestate}
|\Psi\ra=\sum\limits_{i_{1},i_{2},\cdots
i_{N}=1}^{d}a_{i_{1},i_{2},\cdots i_{N}}|i_{1},i_{2},\cdots
i_{N}\ra,\quad a_{i_{1},i_{2},\cdots i_{N}}\in \Cb.
\end{eqnarray}

Let $\alpha$ and $\alpha^{'}$ (resp.$\beta$ and $\beta^{'}$) be
subsets of the subindices of $a$, associated to the same sub Hilbert
spaces but with different summing indices. $\alpha$ (or
$\alpha^{'}$) and $\beta$ (or $\beta^{'}$) span the whole space of the
given sub-indix of $a$. The generalized concurrence of $|\Psi\ra$ is
then given by \cite{anote}
\begin{eqnarray}\label{def}
C_{d}^{N}(|\Psi\ra)=\sqrt{\frac{d}{2m(d-1)}\sum\limits_{p}
\sum\limits_{\{\alpha,\alpha^{'},\beta,\beta^{'}\}}^{d}
|a_{\alpha\beta}a_{\alpha^{'}\beta^{'}}-a_{\alpha\beta^{'}}a_{\alpha^{'}\beta}|^{2}},
\end{eqnarray}
where $m=2^{N-1}-1$, $\sum\limits_{p}$ stands for the summation
over all possible combinations of the indices of $\alpha$ and $\beta$.

For a mixed state $\rho$,
\begin{eqnarray}\label{rho}
\rho=\sum_{i}p_{i}|\psi_{i}\ra\la\psi_{i}|,\quad p_{i}\geq
0,\quad\sum_{i}p_{i}=1,
\end{eqnarray}
the concurrence is defined by the convex-roof:
\begin{eqnarray}\label{conrho}
C(\rho)=\min\sum_{i}p_{i}C(|\psi_{i}\ra),
\end{eqnarray}
minimized over all possible pure state decompositions.

\section{\bf Lower bound of the concurrence of a multipartite quantum state}
We first consider tripartite case. A general pure state on $H\otimes
H\otimes H$ is of the form
\begin{eqnarray}\label{pure}
|\Psi\ra=\sum\limits_{i,j,k=1}^{d}a_{ijk}|ijk\ra,\quad a_{ijk}\in
\Cb,\quad \sum\limits_{i,j,k=1}^{d}a_{ijk}a_{ijk}^{*}=1
\end{eqnarray}
with
\begin{eqnarray}\label{def1}
C_{d}^{3}(|\Psi\ra)=\sqrt{\frac{d}{6(d-1)}\sum(|a_{ijk}a_{pqm}-a_{ijm}a_{pqk}|^{2}
+|a_{ijk}a_{pqm}-a_{iqk}a_{pjm}|^{2}+|a_{ijk}a_{pqm}-a_{pjk}a_{iqm}|^{2})}
\end{eqnarray}
or equivalently
\begin{eqnarray}\label{def12}
C_{d}^{3}(|\Psi\ra)=\sqrt{\frac{d}{6(d-1)}(3-(Tr\rho_{1}^{2}+Tr\rho_{2}^{2}+Tr\rho_{3}^{2}))},
\end{eqnarray}
where $\rho_{1}=Tr_{23}(\rho), \rho_{2}=Tr_{13}(\rho),
\rho_{3}=Tr_{12}(\rho)$ are the reduced density matrices of
$\rho=|\Psi\ra\la\Psi|$.

Define
\begin{eqnarray}
C_{\alpha\beta}^{12|3}(|\Psi\ra)&=&|a_{ijk}a_{pqm}-a_{ijm}a_{pqk}|,\quad
C_{\alpha\beta}^{13|2}(|\Psi\ra)=|a_{ijk}a_{pqm}-a_{iqk}a_{pjm}|,\nonumber\\
C_{\alpha\beta}^{23|1}(|\Psi\ra)&=&|a_{ijk}a_{pqm}-a_{pjk}a_{iqm}|,
\end{eqnarray}
where $\alpha$ and $\beta$ of $C_{\alpha\beta}^{12|3}$ (resp. $C_{\alpha\beta}^{13|2}$ resp. $C_{\alpha\beta}^{23|1}$)
stand for the sub-indices of $a$ associated with the subspaces $1,2$ and $3$ (resp. $1,3$ and $2$ resp. $2,3$ and $1$).
Let $L^{i_{1}i_{2}\cdots i_{N}}$ denote the generators of
group $SO(d_{i_{1}}d_{i_{2}}\cdots d_{i_{N}})$ associated to the
subsystems $i_{1},i_{2},\cdots, i_{N}$. Then for a tripartite pure
state ($\ref{pure}$), one has
\begin{eqnarray}\label{con}
C_{d}^{3}(|\Psi\ra)&=&\sqrt{\frac{d}{6(d-1)}
\sum_{\alpha}^{\frac{d^{2}(d^{2}-1)}{2}}\sum_{\beta}^{\frac{d(d-1)}{2}}
[({C_{\alpha\beta}^{12|3}(|\Psi\ra)})^{2}
+({C_{\alpha\beta}^{13|2}}(|\Psi\ra))^{2}+({C_{\alpha\beta}^{23|1}(|\Psi\ra)})^{2}]}\nonumber\\
&=&\sqrt{\frac{d}{6(d-1)} \sum_{\alpha\beta}
[(|\la\Psi|S_{\alpha\beta}^{12|3}|\Psi^{*}\ra|)^{2}
+(|\la\Psi|S_{\alpha\beta}^{13|2}|\Psi^{*}\ra|)^{2}+(|\la\Psi|S_{\alpha\beta}^{23|1}|\Psi^{*}\ra|)^{2}]},
\end{eqnarray}
where $S_{\alpha\beta}^{12|3}=(L_{\alpha}^{12}\otimes
L^{3}_{\beta})$, $S_{\alpha\beta}^{13|2}=(L_{\alpha}^{13}\otimes
L^{2}_{\beta})$ and $S_{\alpha\beta}^{23|1}=(L_{\beta}^{1}\otimes
L^{23}_{\alpha})$.

{\bf{Theorem 1:}} For an arbitrary mixed state $(\ref{rho})$ in
$H\otimes H\otimes H$, the concurrence $C(\rho)$ satisfies
\begin{eqnarray}\label{11}
\tau_{3}(\rho)\equiv\frac{d}{6(d-1)}\sum_{\alpha}^{\frac{d^{2}(d^{2}-1)}{2}}\sum_{\beta}^{\frac{d(d-1)}{2}}
[({C_{\alpha\beta}^{12|3}(\rho)})^{2}
+({C_{\alpha\beta}^{13|2}(\rho)})^{2}+({C_{\alpha\beta}^{23|1}(\rho)})^{2}]\leq C^{2}(\rho),
\end{eqnarray}
where $\tau_{3}(\rho)$ is a lower bound of $C(\rho)$,
\begin{eqnarray}
C_{\alpha\beta}^{12|3}(\rho)=\max\{0,\lambda(1)_{\alpha\beta}^{12|3}-\lambda(2)_{\alpha\beta}^{12|3}
-\lambda(3)_{\alpha\beta}^{12|3}-\lambda(4)_{\alpha\beta}^{12|3}\},
\end{eqnarray}
$\lambda(1)_{\alpha\beta}^{12|3}, \lambda(2)_{\alpha\beta}^{12|3},
\lambda(3)_{\alpha\beta}^{12|3}, \lambda(4)_{\alpha\beta}^{12|3}$
are the square roots of the four nonzero eigenvalues, in decreasing
order, of the non-Hermitian matrix
$\rho\widetilde{\rho}_{\alpha\beta}^{12|3}$ with
$\widetilde{\rho}_{\alpha\beta}^{12|3}=S_{\alpha\beta}^{12|3}\rho^{*}S_{\alpha\beta}^{12|3}$.
$C_{\alpha\beta}^{13|2}(\rho)$ and $C_{\alpha\beta}^{23|1}(\rho)$
are defined in a similar way to $C_{\alpha\beta}^{12|3}(\rho)$.

{\bf{Proof:}} Set $|\xi_{i}\ra=\sqrt{p_{i}}|\psi_{i}\ra$,
$x_{\alpha\beta}^{i}=|\la\xi_{i}|S_{\alpha\beta}^{12|3}|\xi_{i}^{*}\ra|$,
$y_{\alpha\beta}^{i}=|\la\xi_{i}|S_{\alpha\beta}^{13|2}|\xi_{i}^{*}\ra|$
and
$z_{\alpha\beta}^{i}=|\la\xi_{i}|S_{\alpha\beta}^{1|23}|\xi_{i}^{*}\ra|$.
We have, from Minkowski inequality
\begin{eqnarray}\label{14}
C(\rho)&=&\min\sum_{i}\sqrt{\frac{d}{6(d-1)} \sum_{\alpha\beta}
\left[(x_{\alpha\beta}^{i})^{2} +(y_{\alpha\beta}^{i})^{2}
+(z_{\alpha\beta}^{i})^{2}\right]}\nonumber\\
&\geq&\min\sqrt{\frac{d}{6(d-1)}
\sum_{\alpha\beta}\left(\sum_{i}[(x_{\alpha\beta}^{i})^{2}
+(y_{\alpha\beta}^{i})^{2}+(z_{\alpha\beta}^{i})^{2}]^{\frac{1}{2}}\right)^{2}}.\nonumber
\end{eqnarray}

Noting that for nonnegative real variables $x_{\alpha}$,
$y_{\alpha}$, $z_{\alpha}$ and given
$X=\sum\limits_{\alpha=1}^{N}x_{\alpha}$,
$Y=\sum\limits_{\alpha=1}^{N}Y_{\alpha}$ and
$Z=\sum\limits_{\alpha=1}^{N}z_{\alpha}$, by using Lagrange
multipliers one obtain that the following inequality holds,
\begin{eqnarray}\label{inequality}
\sum\limits_{\alpha=1}^{N}(x_{\alpha}^{2}+y_{\alpha}^{2}+z_{\alpha}^{2})^{\frac{1}{2}}\geq
(X^{2}+Y^{2}+Z^{2})^{\frac{1}{2}}.
\end{eqnarray}
Therefore we have
\begin{eqnarray}\label{14p}
C(\rho)&\geq&\min\sqrt{\frac{d}{6(d-1)}
\sum_{\alpha\beta}[(\sum_{i}x_{\alpha\beta}^{i})^{2}
+(\sum_{i}y_{\alpha\beta}^{i})^{2}+(\sum_{i}z_{\alpha\beta}^{i})^{2}]}\nonumber\\
&\geq&\sqrt{\frac{d}{6(d-1)}
\sum_{\alpha\beta}[(\min\sum_{i}x_{\alpha\beta}^{i})^{2}
+(\min\sum_{i}y_{\alpha\beta}^{i})^{2}+(\min\sum_{i}z_{\alpha\beta}^{i})^{2}]}.
\end{eqnarray}

The values of
$C_{\alpha\beta}^{12|3}(\rho)\equiv\min\sum\limits_{i}x_{\alpha\beta}^{i}$,
$C_{\alpha\beta}^{13|2}(\rho)\equiv\min\sum\limits_{i}y_{\alpha\beta}^{i}$
and
$C_{\alpha\beta}^{23|1}(\rho)\equiv\min\sum\limits_{i}z_{\alpha\beta}^{i}$
can be calculated by using the similar procedure in \cite{wotters}. Here we
compute the value of $C_{\alpha\beta}^{12|3}(\rho)$ in detail. The values of
$C_{\alpha\beta}^{13|2}(\rho)$ and $C_{\alpha\beta}^{23|1}(\rho)$
can be obtained analogously.

Let $\lambda_{i}$ and $|\chi_{i}\ra$ be eigenvalues and eigenvectors
of $\rho$ respectively. Any decomposition of $\rho$ can be obtained
from a unitary $d^{3}\times d^{3}$ matrix $V_{ij}$,
$|\xi_{j}\ra=\sum\limits_{i=1}^{d^{3}}V^{*}_{ij}(\sqrt{\lambda_{i}}|\chi_{i}\ra)$.
Therefore one has
$\la\xi_{i}|S^{12|3}_{\alpha\beta}|\xi_{j}^{*}\ra=(VY_{\alpha\beta}V^{T})_{ij}$,
where the matrix $Y_{\alpha\beta}$ is defined by
$(Y_{\alpha\beta})_{ij}=\la\chi_{i}|S^{12|3}_{\alpha\beta}|\chi_{j}^{*}\ra$.
Namely
$C_{\alpha\beta}^{12|3}(\rho)=\min\sum_{i}|[VY_{\alpha\beta}V^{T}]_{ii}|$,
which has an analytical expression \cite{wotters},
$C_{\alpha\beta}^{12|3}(\rho)=\max\{0,\lambda(1)_{\alpha\beta}^{12|3}
-\sum_{j>1}\lambda(j)_{\alpha\beta}^{12|3}\}$, where
$\lambda^{12|3}_{\alpha\beta}(k)$ are the square roots of the
eigenvalues of the positive Hermitian matrix
$Y_{\alpha\beta}Y_{\alpha\beta}^{\dag}$, or equivalently the
non-Hermitian matrix $\rho\widetilde{\rho}_{\alpha\beta}$, in
decreasing order. Here as the matrix $S_{\alpha\beta}^{12|3}$ has
$d^{2}-4$ rows and $d^{2}-4$ columns that are identically zero, the
matrix $\rho\widetilde{\rho}_{\alpha\beta}$ has a rank no greater
than 4, i.e., $\lambda_{\alpha\beta}^{12|3}(j)=0$ for $j\geq 5$.
From Eq.($\ref{14p}$) we have Eq.($\ref{11}$).$\hfill\Box$

Theorem 1 can be directly generalized to arbitrary multipartite
case.

{\bf{Theorem 2:}} For an arbitrary $N$-partite state $\rho\in
{{H}}\otimes {H}\otimes\cdots\otimes {H}$, the concurrence defined
in ($\ref{conrho}$) satisfies:
\begin{eqnarray}
\tau_{N}(\rho)\equiv\frac{d}{2m(d-1)}\sum_{p}\sum_{\alpha\beta}(C_{\alpha\beta}^{p}(\rho))^{2}\leq
C^{2}(\rho),
\end{eqnarray}
where $\tau_{N}(\rho)$ is the lower bound of $C(\rho)$,
$\sum\limits_{p}$ stands for the summation over all possible
combinations of the indices of $\alpha,\beta$,
$C_{\alpha\beta}^{p}(\rho)=\max\{0,
\lambda(1)_{\alpha\beta}^{p}-\lambda(2)_{\alpha\beta}^{p}
-\lambda(3)_{\alpha\beta}^{p}-\lambda(4)_{\alpha\beta}^{p}\}$,
$\lambda(i)_{\alpha\beta}^{p}$, $i=1, 2, 3, 4$, are the square roots
of the four nonzero eigenvalues, in decreasing order, of the
non-Hermitian matrix $\rho\widetilde{\rho}_{\alpha\beta}^{p}$ where
$\widetilde{\rho}_{\alpha\beta}^{p}=S_{\alpha\beta}^{p}\rho^{*}S_{\alpha\beta}^{p}$.

\section{\bf The lower bound and separability}
An N-partite quantum state $\rho$ is fully separable if and only if
there exist $p_{i}$ with $p_{i}\geq0, \sum\limits_{i}p_{i}=1$ and
pure states $\rho_{i}^{j}=|\psi_{i}^{j}\ra\la\psi_{i}^{j}|$ such
that
\begin{eqnarray}
\rho=\sum_{i}p_{i}\rho_{i}^{1}\otimes\rho_{i}^{2}\otimes\cdots\otimes\rho_{i}^{N}.
\end{eqnarray}

It is easily verified that for a fully separable multipartite state
$\rho$, $\tau_{N}(\rho)=0$. Thus $\tau_{N}(\rho)>0$ indicates that
there must be some kinds of entanglement inside the quantum state,
which shows that the lower bound $\tau_{N}(\rho)$ can be used to
recognize entanglement.

As an example we consider a tripartite quantum state \cite{acin},
$\rho=\frac{1-p}{8}I_{8}+p|W\ra\la W|$, where $I_{8}$ is the
$8\times8$ identity matrix, and
$|W\ra=\frac{1}{\sqrt{3}}(|100\ra+|010\ra+|001\ra)$ is the
tripartite W-state. Select an entanglement witness operator to be
${\mathcal {W}}=\frac{1}{2}I_{8}-|GHZ\ra\la GHZ|$, where
$|GHZ\ra=\frac{1}{\sqrt{2}}(|000\ra+|111\ra)$ to be the tripartite
GHZ-state. By computing $Tr\{{\mathcal {W}}\rho\}<0$ the
entanglement of $\rho$ is detected for $\frac{3}{5} < p \leq 1$ in
\cite{acin}. In \cite{hassan} the authors have obtained the
generalized correlation matrix criterion which says if an N-qubit
quantum state is fully separable then the inequality $||{\mathcal
{T}}^{N}||_{KF}\leq 1$ must hold, where $||{\mathcal
{T}}^{N}||_{KF}=\max\{||{\mathcal {T}}_{n}^{N}||_{KF}\}$, ${\mathcal
{T}}_{n}^{N}$ is a kind of matrix unfold of
$t_{\alpha_{1}\alpha_{2}\cdots\alpha_{N}}$ defined by
$t_{\alpha_{1}\alpha_{2}\cdots\alpha_{N}}=Tr\{\rho\sigma_{\alpha_{1}}^{(1)}
\sigma_{\alpha_{2}}^{(2)}\cdots \sigma_{\alpha_{N}}^{(N)}\}$ and
$\sigma_{\alpha_{i}}^{(i)}$ stands for the pauli matrix. Now using
the generalized correlation matrix criterion the entanglement of
$\rho$ is detected for $0.3068 < p \leq 1$. From our theorem, we
have that the lower bound $\tau_{3}(\rho)>0$ for $0.2727 < p \leq
1$. Therefore our bound detects entanglement better than these two
criteria in this case. If we replace W with GHZ state in $\rho$, the
criterion in \cite{hassan} detects the entanglement of $\rho$ for
$0.35355 < p \leq 1$, while $\tau_{3}(\rho)$ detects, again better,
the entanglement for $0.2 < p \leq 1$.

Nevertheless for PPT states $\rho$, we have $\tau_{3}(\rho)=0$,
which can be seen in the following way. A density matrix $\rho$ is
called PPT if the partial transposition of $\rho$ over any
subsystem(s) is still positive. Let $\rho^{T_{i}}$ denote the
partial transposition with respect to the $i$-th subsystem. Assume
that there is a PPT state $\rho$ with $\tau(\rho)>0$. Then at least
one term in ($\ref{11}$), say
$C_{\alpha_{0}\beta_{0}}^{12|3}(\rho)$, is not zero. Define
$\rho_{\alpha_{0}\beta_{0}}=L_{\alpha_{0}}^{12}\otimes
L_{\beta_{0}}^{3}\rho (L_{\alpha_{0}}^{12}\otimes
L_{\beta_{0}}^{3})^{\dag}$. By using the PPT property of $\rho$, we
have:
\begin{eqnarray}\label{19}
\rho_{\alpha_{0}\beta_{0}}^{T_{3}}=L_{\alpha_{0}}^{12}\otimes
(L_{\beta_{0}}^{3})^{*}\rho^{T_{3}}
(L_{\alpha_{0}}^{12})^{\dag}\otimes (L_{\beta_{0}}^{3})^{T}\geq 0.
\end{eqnarray}
Noting that both $L_{\alpha_{0}}^{12}$ and $ L_{\beta_{0}}^{3}$ are
projectors to two-dimensional subsystems,
$\rho_{\alpha_{0}\beta_{0}}$ can be considered as a $4\times 4$
density matrix. While a PPT $4\times 4$ density matrix
$\rho_{\alpha_{0}\beta_{0}}$ must be a separable state,
which contradicts with $C_{\alpha_{0}\beta_{0}}^{12|3}(\rho)\neq 0$.

\section{\bf Comparison with the lower bound of the bipartite
concurrence}
The lower bound $\tau_{2}$ of concurrence for bipartite
states has been obtained in \cite{ou}. For a bipartite quantum state
$\rho$ in $H\otimes H$, the concurrence $C(\rho)$ satisfies
\begin{eqnarray}\label{ou}
\tau_{2}(\rho)\equiv\frac{d}{2(d-1)}\sum_{m,n=1}^{\frac{d(d-1)}{2}}C_{mn}^{2}(\rho)\leq
C^{2}(\rho),
\end{eqnarray}
where $C_{mn}^{2}(\rho)=\max\{0,\lambda_{mn}(1)-\lambda_{mn}(2)
-\lambda_{mn}(3)-\lambda_{mn}(4)\}$ with $\lambda_{mn}(1), ...,
\lambda_{mn}(4)$ being the square roots of the four nonzero
eigenvalues, in decreasing order, of the non-Hermitian matrix
$\rho\widetilde{\rho}_{mn}$ with
$\widetilde{\rho}_{mn}=(L_{m}\otimes L_{n})\rho^{*}(L_{m}\otimes
L_{n})$, $L_{m}$ and $L_{n}$ being the generators of $SO(d)$.
$\tau_{3}$ is basically different from $\tau_{2}$ as $\tau_{3}$
characterizes also genuine tripartite entanglement that can not be
described by bipartite decompositions. Nevertheless, there are
interesting relations between them.

{\bf{Theorem 3:}} For any pure tripartite state ($\ref{pure}$), the
following inequality holds:
\begin{eqnarray}
\tau_{2}(\rho_{12})+\tau_{2}(\rho_{13})+\tau_{2}(\rho_{23}) \leq
3\tau_{3}(\rho),
\end{eqnarray}
where $\tau_{2}$ is the lower bound of bipartite concurrence
($\ref{ou}$), $\tau_{3}$ is the lower bound of
tripartite concurrence ($\ref{11}$) and $\rho_{12}=Tr_{3}(\rho)$,
$\rho_{13}=Tr_{2}(\rho)$, $\rho_{23}=Tr_{1}(\rho)$,
$\rho=|\Psi\ra_{123}\la\Psi|$.

{\bf{Proof:}} Since
$C_{\alpha\beta}^{2}\leq(\lambda_{\alpha\beta}(1))^{2}
\leq\sum_{i=1}^{4}(\lambda_{\alpha\beta}(i))^{2}=Tr(\rho
\widetilde{\rho}_{\alpha\beta})$ for $\rho=\rho_{12}$,
$\rho=\rho_{13}$ and $\rho=\rho_{23}$, we have
\begin{eqnarray}
&&\tau_{2}(\rho_{12})+\tau_{2}(\rho_{13})+\tau_{2}(\rho_{23})
\nonumber\\
&\leq&
\frac{d}{2(d-1)}(\sum_{\alpha,\beta=1}^{\frac{d(d-1)}{2}}Tr(\rho_{12}
(\widetilde{\rho}_{12})_{\alpha\beta})+\sum_{\alpha,\beta=1}^{\frac{d(d-1)}{2}}Tr(\rho_{13}
(\widetilde{\rho}_{13})_{\alpha\beta})+\sum_{\alpha,\beta=1}^{\frac{d(d-1)}{2}}Tr(\rho_{23}
(\widetilde{\rho}_{23})_{\alpha\beta})) \nonumber\\
&=&\frac{d}{2(d-1)}(3-Tr\rho_{1}^{2}-Tr\rho_{2}^{2}-Tr\rho_{3}^{2})=3C^{2}(\rho)=3\tau_{3}(\rho),
\end{eqnarray}
where we have used the similar analysis in \cite{ou,ckw} to obtain
the equality $\sum\limits_{\alpha,\beta}Tr(\rho_{12}
(\widetilde{\rho}_{12})_{\alpha\beta})=1-Tr\rho_{1}^{2}-Tr\rho_{2}^{2}+Tr\rho_{3}^{2}$,
$\sum\limits_{\alpha,\beta}Tr(\rho_{13}
(\widetilde{\rho}_{13})_{\alpha\beta})=1-Tr\rho_{1}^{2}+Tr\rho_{2}^{2}-Tr\rho_{3}^{2}$,
$\sum\limits_{\alpha,\beta}Tr(\rho_{23}
(\widetilde{\rho}_{23})_{\alpha\beta})=1+Tr\rho_{1}^{2}-Tr\rho_{2}^{2}-Tr\rho_{3}^{2}$.
The last equality is due to that $\rho$ is a pure state.\hfill$\Box$

In fact, the bipartite entanglement inside a tripartite state
is useful for distilling maximally entangled states.
Assume that there are two of the qualities $\{\tau(\rho_{12}),
\tau(\rho_{13}), \tau(\rho_{23})\}$ larger than zero, say
$\tau(\rho_{12})>0$ and $\tau(\rho_{13})>0$. According to \cite{ou},
one can distill two maximal entangled states $|\psi_{12}\ra$ and
$|\psi_{13}\ra$ which belong to ${\mathcal {H}}_{1}\otimes{\mathcal
{H}}_{2}$ and ${\mathcal {H}}_{1}\otimes{\mathcal {H}}_{3}$
respectively. In terms of the result in \cite{zukowski}, one
can use them to produce a GHZ state.

\section{\bf Conclusions}

We have studied the concurrence for multipartite quantum states and
derived an explicit lower bound of the concurrence.
This bound can be also served as separability
criterion. It detects entanglement of some states better than some
separability criteria. For tripartite PPT states the lower bound is
zero. The bound also gives sufficient conditions for distilling GHZ
states from tripartite states. Moreover it has been shown that there
is an interesting relation, similar to the monogamy inequalities and
tangle \cite{tangle}, between the lower bound of the concurrence
$\tau_2$ for bipartite states and $\tau_3$ for tripartite states. In
addition, our results can be easily generalized to the situation
that all the subsystems have different dimensions. By simply neglect
the coefficient related to the dimensions, $\frac{d}{2m(d-1)}$, in
the concurrence defined in ($\ref{def}$), similar results of theorem
2 and 3 hold for systems with different dimensions of subsystems.

\bigskip
\noindent{\bf Acknowledgments}\, This work is supported by the NSFC
10675086, KZ200810028013, and NKBRPC(2004CB318000).


\smallskip
\begin{thebibliography}{99}

\bibitem{nielsen} Nielsen M A, Chuang I L. Quantum Computation and Quantum
Information. Cambridge: Cambridge University Press, (2000).

\bibitem{eof} C. H. Bennett, D. P. DiVincenzo and J. A. Smolin, et al. Phys. Rev. A
54, 3824(1996).\\
 M. B. Plenio and S. Virmani, Quant. Inf. Comp. 7, 1(2007).

\bibitem{concurrence} A. Uhlmann  Phys. Rev. A 62 032307(2000);\\
P. Rungta, V. Bu$\breve{z}$ek, and C. M. Caves, et al. Phys. Rev. A 64, 042315(2001).

\bibitem{anote}
S. Albeverio and S. M. Fei, J. Opt. B: Quantum Semiclass Opt, 3,
223-227(2001).

\bibitem{wotters} W. K. Wootters, Phys. Rev. Lett. 80, 2245 (1998).

\bibitem{Terhal-Voll2000}  Terhal B M,  Vollbrecht K G H,  Phys Rev
Lett, 85, 2625-2628(2000).

\bibitem{fjlw} S.M. Fei, J. Jost, X.Q. Li-Jost and G.F. Wang, Phys. Lett. A 310, 333-338(2003).

\bibitem{fl} S.M. Fei and X.Q. Li-Jost, Rep. Math. Phys. 53, 195-210(2004).

\bibitem{fwz} S.M. Fei, Z.X. Wang and H. Zhao, Phys. Lett. A 329, 414-419(2004).

\bibitem{Rungta03} P. Rungta and C.M. Caves, Phys Rev A 67, 012307(2003).

\bibitem{167902} F. Mintert, M. Kus, A. Buchleitner, Phys. Rev. Lett. 92, 167902(2004).

\bibitem{Chen-Albeverio-Fei1} K. Chen, S. Albeverio and S.M. Fei, Phys. Rev. Lett. 95, 210501(2005).

\bibitem{chen} K. Chen, S. Albeverio, and S. M. Fei, Phys. Rev. Lett. 95,
040504(2005).

\bibitem{breuer} H.P. Breuer, J. Phys. A 39, 11847(2006).

\bibitem{breuerprl} H.P. Breuer, Phys. Rev. Lett. 97, 080501(2006).

\bibitem{vicente} J. I. de Vicente, Phys. Rev. A 75, 052320(2007).

\bibitem{zhang} C. J. Zhang, Y. S. Zhang, and S. Zhang, et al. Phys. Rev. A 76, 012334(2007).

\bibitem{gao} X. H. Gao, S. M. Fei and K. Wu, Phys. Rev. A 74, 050303(R) (2006).

\bibitem{edward} E. Gerjuoy, Phys. Rev. A 67, 052308(2003).

\bibitem{ou} Y. C. Ou, H. Fan and S. M. Fei, Phys. Rev. A 78, 012311(2008).

\bibitem{acin} A. Ac$\acute{i}$n, D. Bruss, M. Lewenstein, and A. Sanpera, Phys. Rev. Lett.
87, 040401(2001).

\bibitem{hassan} A. S. M. Hassan and P. S. Joag, Quant. Inf. Comp. 8, 0773-0790(2008).

\bibitem{charles} C. H. Bennet, D. P. Divincenzo, Tal Mor, P. W.
Shor, J. A. Smolin and B. M. Terhal, Phys. Rev. Lett. 82, 2881(1999).

\bibitem{pittenger} A. O. Pittenger, and M. H. Rubin, Phys, Rev. A 62, 042306 (2000).

\bibitem{ckw} V. Coffman, J. Kundu, and W. K. Wootters, Phys. Rev. A 61, 052306 (2000).

\bibitem{zukowski} M. Zukowski, A. Zeilinger, M. A. Horne and A. K. Ekert, Phys. Rev. Lett. 71, 4287 (1993).

\bibitem{tangle} V. Coffman, J. Kundu, and W. K. Wootters, Phys. Rev.
A 61, 052306(2000).\\
T. J. Osborne and F. Verstraete, Phys. Rev. Lett. 96, 220503(2006).\\
Y.C. Ou, Phys. Rev. A 75, 034305(2007).\\
W. D\"ur, G. Vidal, J. I. Cirac, Phys. Rev. A 62, 062314(2000).\\
Y.C. Ou, H. Fan, Phys. Rev. A 76, 022320(2007).

\end{thebibliography}
\end{document}